\documentclass[letter,twocolumn]{jpsj2}
\usepackage{color}

\title{%
Finite-Temperature Phase Diagram 
of Quasi-One-Dimensional Molecular Conductors:
Quantum Monte Carlo Study
}

\author{%
Yuichi \textsc{Otsuka}$^{1,2}$%
\thanks{%
E-mail: otsuka@sci.u-hyogo.ac.jp 
}
Hitoshi \textsc{Seo}$^{2}$,
Yukitoshi \textsc{Motome}$^{3}$, and
Takeo \textsc{Kato}$^{4}$%
}

\inst{%
$^{1}$ CREST, Japan Science and Technology Agency, Kawaguchi, Saitama 332-0012\\
$^{2}$ Synchrotron Radiation Research Center, Japan Atomic Energy Agency, SPring-8, Hyogo 679-5148\\
$^{3}$ Department of Applied Physics, University of Tokyo, Hongo Bunkyo-ku, Tokyo 113-8656\\
$^{4}$ Institute for Solid State Physics, University of Tokyo, Kashiwa, Chiba 277-8581%
}

\abst{% 
Finite-temperature phase transitions in
quasi-one-dimensional quarter-filled systems 
are investigated using
the extended Hubbard model with electron-lattice coupling. 
By a quantum Monte Carlo method
combined with interchain mean-field approximation, 
we clarify competing and coexisting behaviors 
among charge ordering, dimer Mott, and spin-Peierls states.
It is pointed out that 
an anharmonicity of lattice distortions 
plays an important role in 
multicritical behaviors.
The results are compared with experimental data 
for quasi-one-dimensional molecular conductors such as DCNQI and TMTTF compounds.
}

\kword{%
quasi-one-dimensional molecular conductors,
extended Hubbard model,
electron-lattice coupling,
charge ordering,
lattice dimerization,
spin-Peierls transition,
SSE Monte Carlo method
}
\begin{document}
\maketitle

%\texttt{[Introduction]}

Quasi-one-dimensional (Q1D) molecular conductors
have been recognized as 
materials suitable for studying the effects of strong correlation.
In these compounds, 
Coulomb interaction 
and enhanced fluctuations due to low dimensionality
give rise to 
keen competitions in charge, spin, and lattice degrees of freedoms. 
As a result, they show various phase transitions 
despite the fact that their noninteracting band structures are often very similar~\cite{Review}. 

A typical case is the family of 
($R_{1}R_{2}$-DCNQI)$_{2}X$ 
($R_{1}$, $R_{2}$: substituents, $X$: monovalent cation). 
The electronic structure is commonly described by 
a Q1D quarter-filled $\pi$-band 
of DCNQI chains~\cite{Miyazaki-Terakura-1996}, 
whereas their physical properties 
depend strongly on $R_{1}$, $R_{2}$, and $X$.
For instance, 
DI-DCNQI$_{2}$Ag shows 
a charge-ordering (CO) phase transition at $T$ = 220 K 
and an antiferromagnetic transition at 5 K~\cite{Hiraki-Kanoda-1996,Hiraki-Kanoda-1998}.
On the other hand, 
DMe-DCNQI$_{2}$Ag exhibits, instead of CO, 
lattice dimerization at $T$ = 100 K, 
which drives the system effectively half-filled,
resulting in a dimer Mott (DM) insulating state.
Furthermore, 
tetramerization occurs at $T$ = 80 K,
which is ascribed to a spin-Peierls (SP) transition~\cite{Moret-1988}.
Such a variety of properties
suggests a subtle balance among different phases
under electron correlation. 

Another good example is found in (TMTTF)$_{2}X$ 
($X$: monovalent anion), 
whose Q1D $\pi$-band of TMTTF chains is quarter-filled 
in terms of holes.~\cite{Jerome-ChemRev2004}
In contrast to the DCNQI systems, 
an intrinsic dimerization exists along the chains from the outset. 
These compounds also exhibit rich phases depending on $X$. 
The CO transition occurs at around 100 K, 
which is followed by a SP transition at a lower temperature ($T$)
for $X$ = PF$_6$ and AsF$_6$ 
or an antiferromagnetic transition for $X$= SbF$_6$~\cite{Chow-2000,Zamborszky,You-2004}. 
For $X$ = AsF$_6$, 
the CO phase is suppressed under applied pressure ($P$),
while the SP state persists up to higher $P$. 
This indicates that 
the CO and SP states are competing in nature~\cite{Zamborszky}.

Theoretically,
correlation effects in such Q1D molecular conductors
have been studied using 
the 1D or Q1D quarter-filled extended Hubbard model 
with on-site and intersite Coulomb interactions, 
mainly focusing on the ground state~\cite{Seo-Merino-Yoshioka-Ogata-2006}. 
Several studies have recently been conducted
to describe their finite-$T$ properties~\cite{Sugiura-2004,Sugiura-2005,Yoshioka-Tsuchiizu-Seo-2006,QTM,Clay-2007,Yoshioka-Tsuchiizu-Seo-2007}. 
In such studies, 
it is crucial to treat 
1D electronic fluctuations properly,
which originate from the strong electron correlation.
This is because development of correlation lengths is strongly
anisotropic along 1D chains, 
which finally results in phase transitions 
through 
interchain interaction or coupling
to the three-dimensional lattice.
Moreover, 
the paramagnetic insulating nature 
in strongly correlated phases such as the CO and DM states 
can be captured 
only when the fluctuation effects are taken into account.
Recently, 
a numerical study including such 1D fluctuation effects was performed~\cite{QTM} 
using the quantum transfer-matrix (QTM) method
combined with the interchain mean-field approximation.
Although competitions among different phases were elucidated,
the low-$T$ region was not fully investigated
because of limitations of the QTM method. 
Most of the intriguing competitions in real materials
appear at much lower $T$; hence,
it is strongly desired to study the low-$T$ region
to understand the experimental results.

\begin{figure}[htbp]
 \begin{center}
  \includegraphics[width=0.22\textwidth,clip]{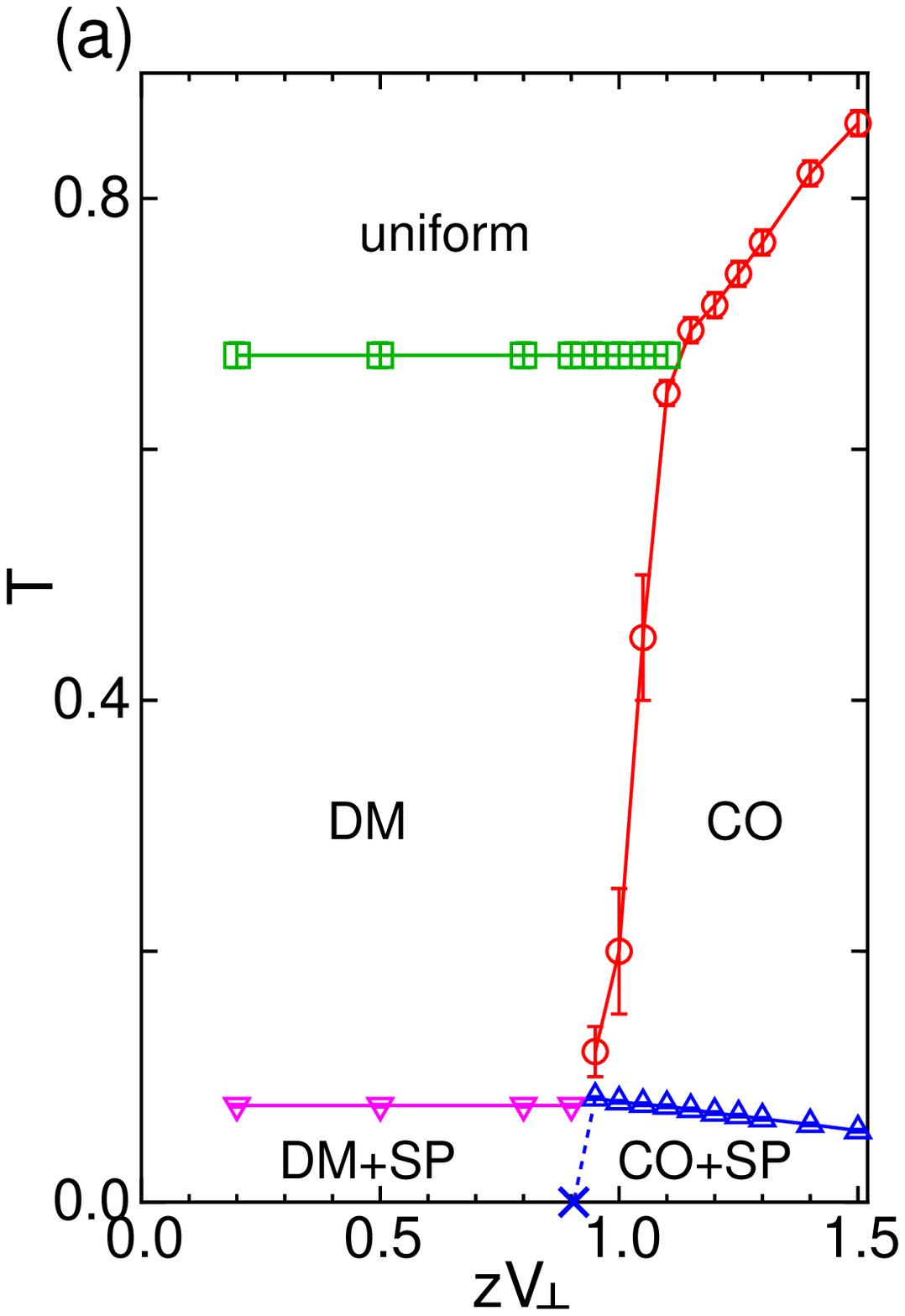}
  \includegraphics[width=0.22\textwidth,clip]{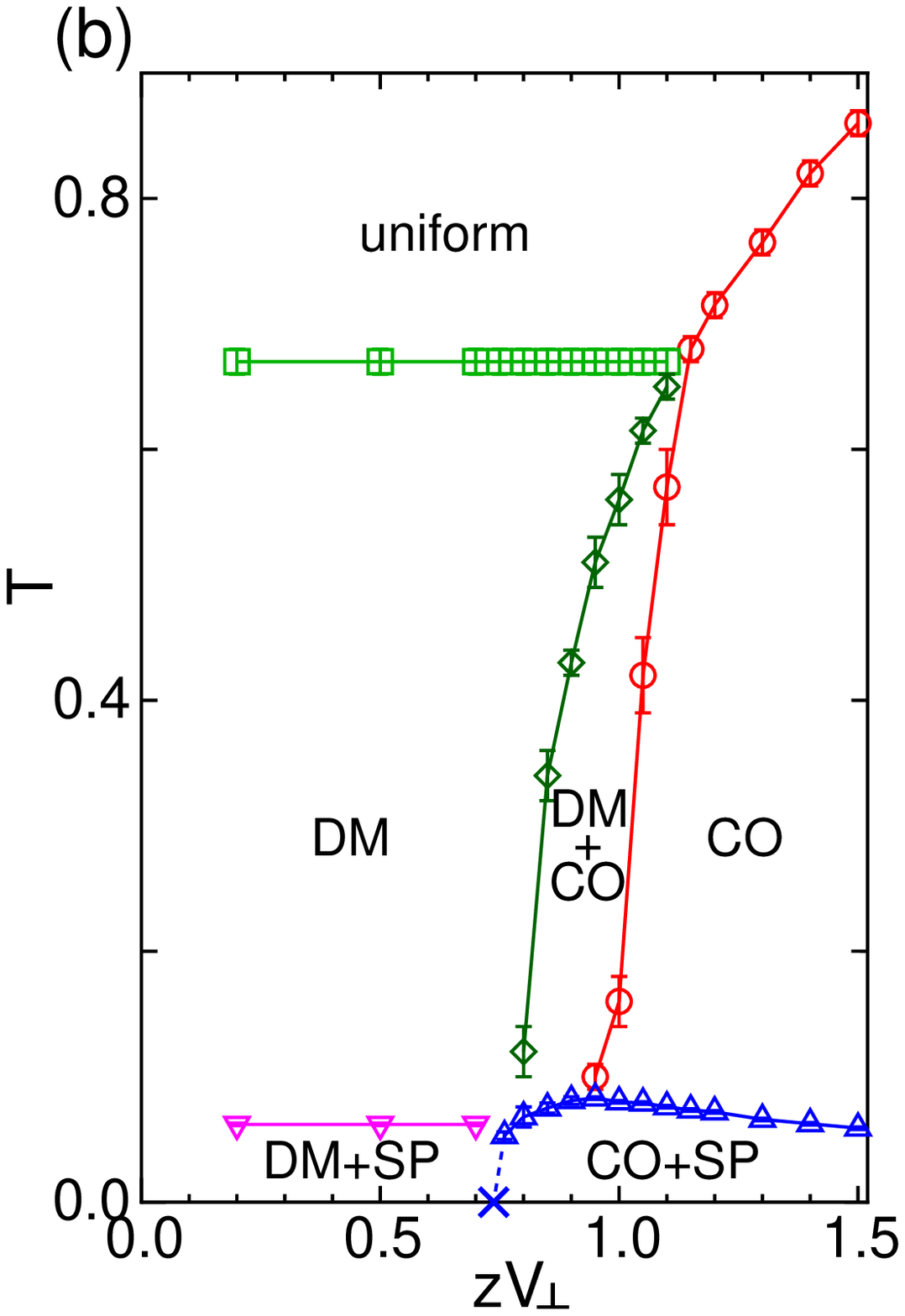}
  \includegraphics[width=0.22\textwidth,clip]{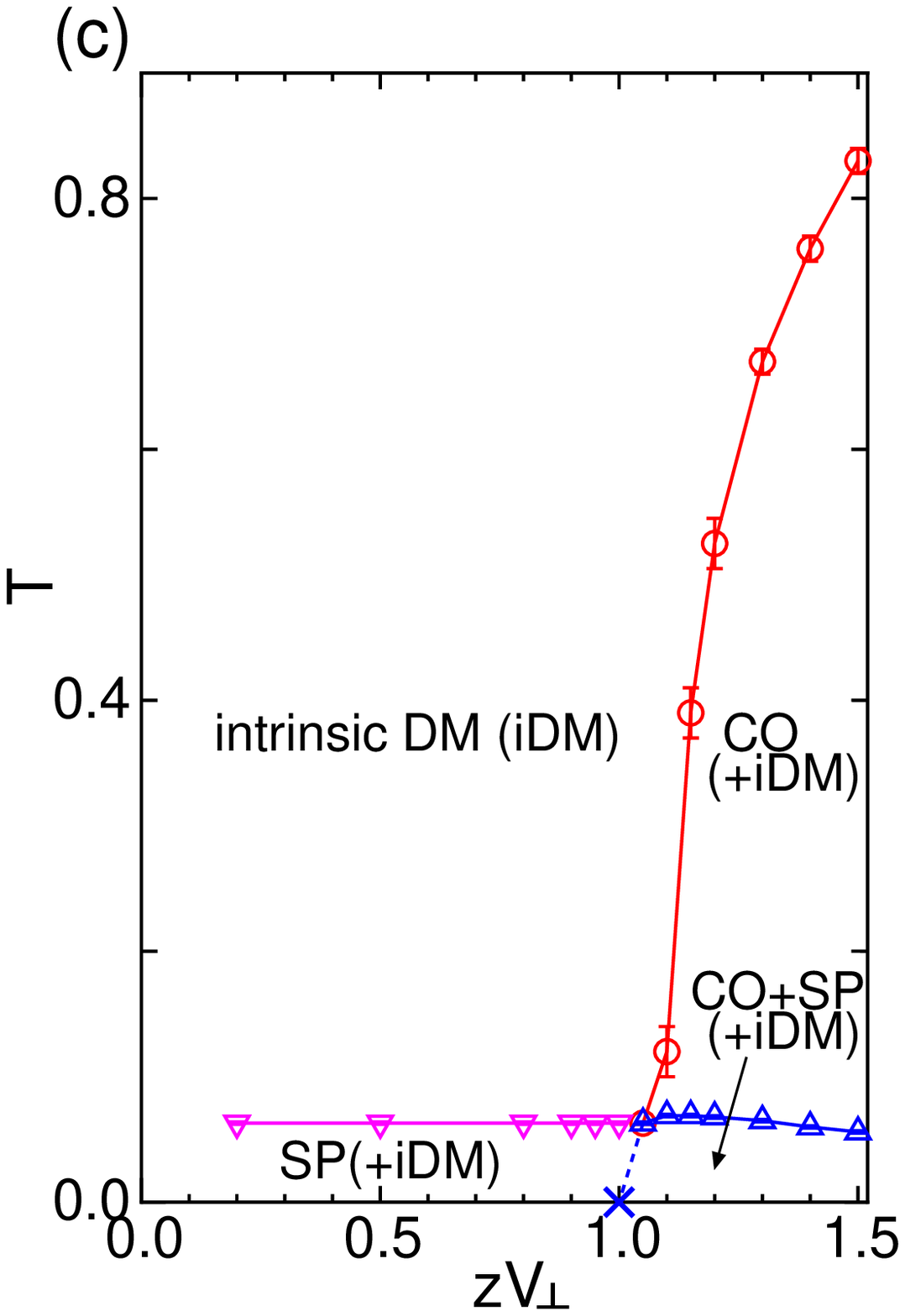}
  \includegraphics[width=0.245\textwidth,clip]{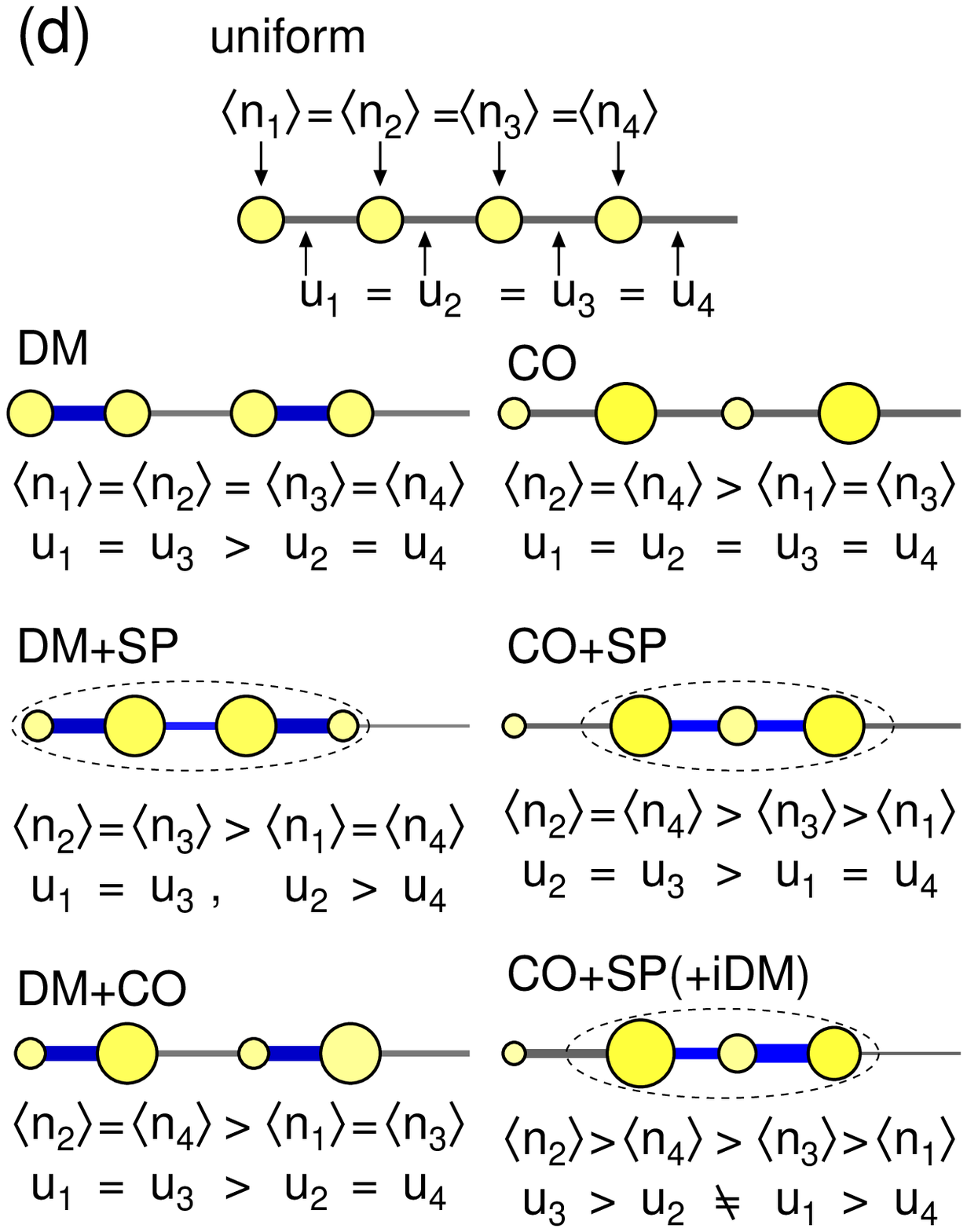}
  \caption{%(Color online):
  Phase diagrams 
  in the plane of $z V_{\perp}$ and $T$
  for 
  $t=1$, $U=6$, $V=2.5$, and $K_{\text{P}}=0.75$;
  (a) $\delta_{\text{d}}=0$    and $K_{\text{P}_{2}}=0$,
  (b) $\delta_{\text{d}}=0$    and $K_{\text{P}_{2}}=0.75$,
  (c) $\delta_{\text{d}}=0.02$ and $K_{\text{P}_{2}}=0$.
  Phase boundaries at $T=0$ (cross symbols) are 
  obtained by the exact diagonalization (Lanczos) method for $N$=12.
  (d) Schematic views of different ordered states.
  Dashed ellipses represent spin-singlets.
  Note that in (c), 
  intrinsic dimerization always exists;
  therefore CO(+iDM) and SP(+iDM) phases 
  respectively have the same patterns as
  DM+CO and  DM+SP  states in (d).
  \label{fig:phase_diagram} 
  }
 \end{center}
\end{figure}

In this paper,
we investigate 
the finite-$T$ properties of Q1D quarter-filled systems,
particularly finite-$T$ phase diagrams 
including various charge and lattice ordered phases,
by a quantum Monte Carlo method 
with interchain mean-field approximation.
This numerical method is reliable enough 
to obtain high-precision results 
down to low $T$ far below that previously studied by QTM.
The main results are summarized in the phase diagrams in Fig.~\ref{fig:phase_diagram},
which show how phase competition and coexistence 
are controlled by
electron correlation and electron-lattice coupling.
We will discuss the results 
in comparison with experiments on the above-mentioned compounds.

\onecolumn
\twocolumn

%\texttt{[Model and Method]}

Our Q1D Hamiltonian is given by 
$\mathcal{H} = \sum_{j} \mathcal{H}_{\text{1D}}^{j} + \mathcal{H}_{\perp}$,
where $\mathcal{H}_{\text{1D}}^{j}$ and $\mathcal{H}_{\perp}$ represent 
the intrachain Hamiltonian of the $j$-th chain
and interchain one, respectively.
The intrachain part is described by 
the extended Hubbard model with the Peierls-type electron-lattice coupling,
whose Hamiltonian reads
\begin{align}
 \mathcal{H}_{\text{1D}}^{j}
 &= 
- \sum_{i, \sigma} 
\{ t + (-1)^i \delta_{\text{d}} \}
\left( 1 + u_{i,j} \right)
\left(
c_{i,j, \sigma}^{\dagger} c_{i+1,j, \sigma}
+
\text{h.c.}
\right)
\nonumber \\%
& +
 \frac{K_{\text{P}    }}{2} \sum_{i} u_{i,j}^{2}
  +
 \frac{K_{\text{P}_{2}}}{2} \sum_{i} u_{i,j}^{4}
\nonumber \\%
& +
 U \sum_{i} n_{i,j \uparrow} n_{i,j\downarrow}
 +
 V \sum_{i} n_{i,j} n_{i+1,j}
 - \mu \sum_{i}  n_{i,j},
\label{eq:model}
\end{align}
where 
the notations are referred to ref.~\citen{QTM}.
The lattice distortions $u_{i,j}$ 
couple to electrons through the transfer integrals, 
and the coupling constant is incorporated in the definition of $u_{i,j}$.
We treat $u_{i,j}$ as classical variables.
The elastic energy is considered up to the fourth order, 
where the term with $K_{\text{P}_{2}}$ represents the anharmonic contribution.
The interchain  part is given by
\begin{equation}
 \mathcal{H}_{\perp} = V_{\perp} \sum_{\langle j, k \rangle} \sum_{i} n_{i,j} n_{i,k},
\end{equation}
where $V_{\perp}$ denotes the interchain Coulomb interaction
and the summation for $\langle j, k \rangle$ runs over nearest-neighbor chains.
In the following calculations, 
we choose the on-site and nearest-neighbor Coulomb interactions
as $U=6$ and $V=2.5$, respectively,~\cite{Kuwabara-Seo-Ogata-2003}
and the elastic constant as $K_{\text{P}}=0.75$, 
in energy unit of $t$.
The chemical potential $\mu$ is controlled so that the system is 
at quarter filling.

We derive an effective 1D model under two approximations.
The first is the interchain mean-field approximation;
by extending eq. (4) in ref.~\citen{QTM} 
to incorporate the CO+SP state in Fig~\ref{fig:phase_diagram}(d),
we assume the mean-field form of $\mathcal{H}_{\perp}$ as
\begin{equation}
 \mathcal{H}_{\perp}^{\text{MF}} = 
\frac{z V_{\perp}}{2} 
\sum_{i} \left\{
\left( 
 \langle n_{i-1} \rangle  + \langle n_{i+1} \rangle 
\right) n_{i}
- \langle n_{i-1} \rangle   \langle n_{i+1} \rangle
\right\}, \label{eq:vperp}
\end{equation}
where $z$ denotes the number of nearest-neighbor chains 
and the chain index $j$ is dropped hereafter.
The second is the adiabatic approximation for the lattice distortions;
we consider the situation where the dynamics of lattices is 
much slower than that of electrons
and fluctuations of electrons play a major role.
The lattice distortions $u_i$  are then determined 
so as to minimize the free energy
under the constraint $\sum_{i} u_{i} = 0$.
Note that we omit the chain index $j$ again, $u_{i,j} = u_i$.
These two approximations give rise to an effective 1D model
$\mathcal{H}_{\text{1D}} + \mathcal{H}_{\perp}^{\text{MF}}$
with self-consistent conditions for $\langle n_i \rangle$ and $u_i$,
which exhibits finite-$T$ phase transitions.
This is suitable for our purpose, 
i.e., to capture the essential features
of the phase transitions governed by 1D electronic fluctuations
which occur at finite $T$ assisted by the additional three-dimensionality.

We solve the effective 1D model 
by a quantum Monte Carlo technique 
called the stochastic-series-expansion (SSE) method~\cite{SSE-1991,SSE-1992}
in the operator-loop-update scheme~\cite{loop,Sengupta}. 
The SSE calculations fully include
thermal and quantum fluctuations of electrons
and give unbiased and high-precision data for the effective 1D model. 
We use the expectation values by SSE 
in the self-consistent equations,
and obtain the inputs for the next SSE calculations in turn.
This cycle is repeated until the convergence is reached
for all $\langle n_i \rangle$ and $u_i$.
We consider symmetry breaking with a four-site period along the chain 
by taking account of tetramerizations in the experiments,
and solve the self-consistent equations 
for eight order parameters:
four each for 
the charge densities $\langle n_{l} \rangle$ 
and the lattice distortions $u_{l}$ ($l=1,2,3,4$).
We have calculated systems with sizes up to $N=64$ sites
and confirmed that finite-size effects are negligible
down to $T=0.02t$.

%\texttt{[Results]}

\begin{figure}[htbp]
 \begin{center}
  \includegraphics[width=0.240\textwidth,clip]{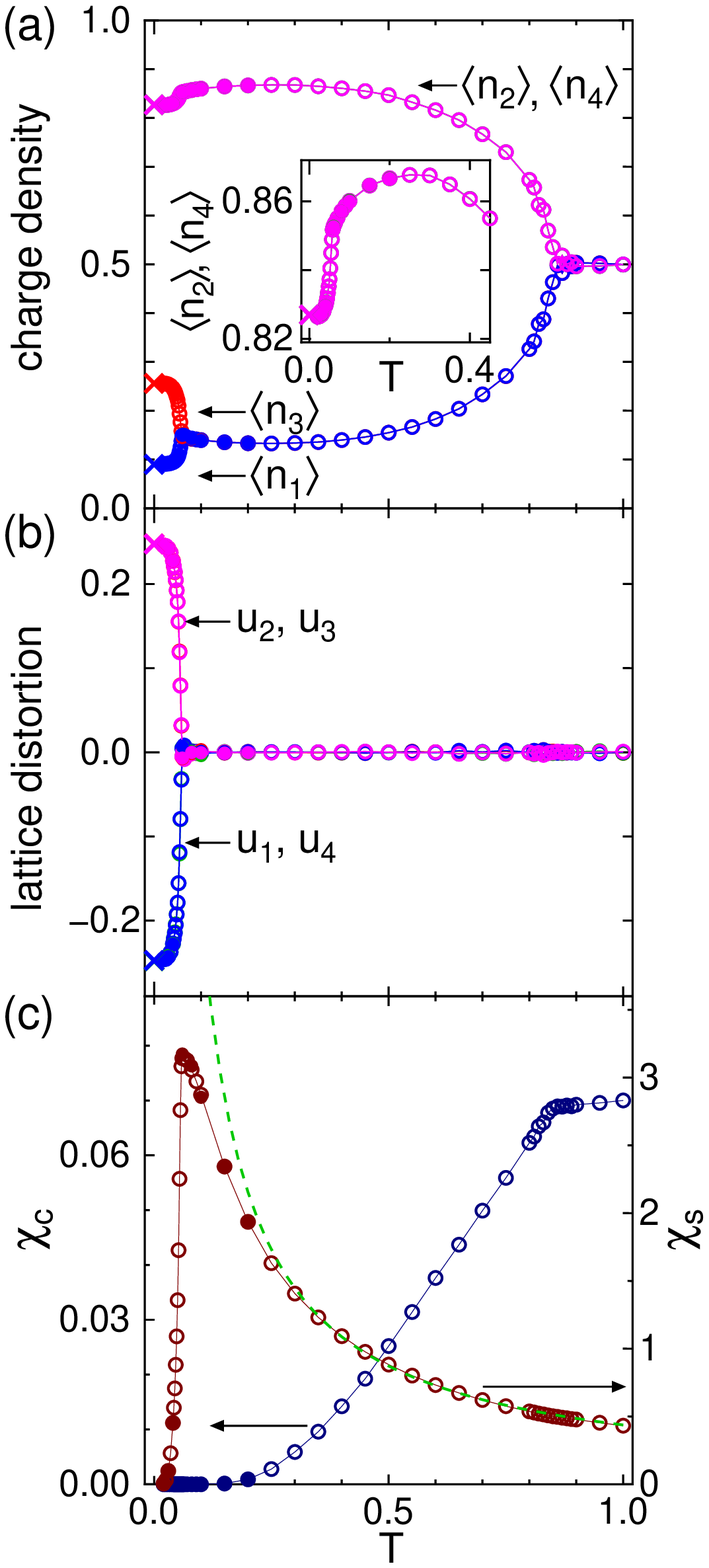}
  \includegraphics[width=0.240\textwidth,clip]{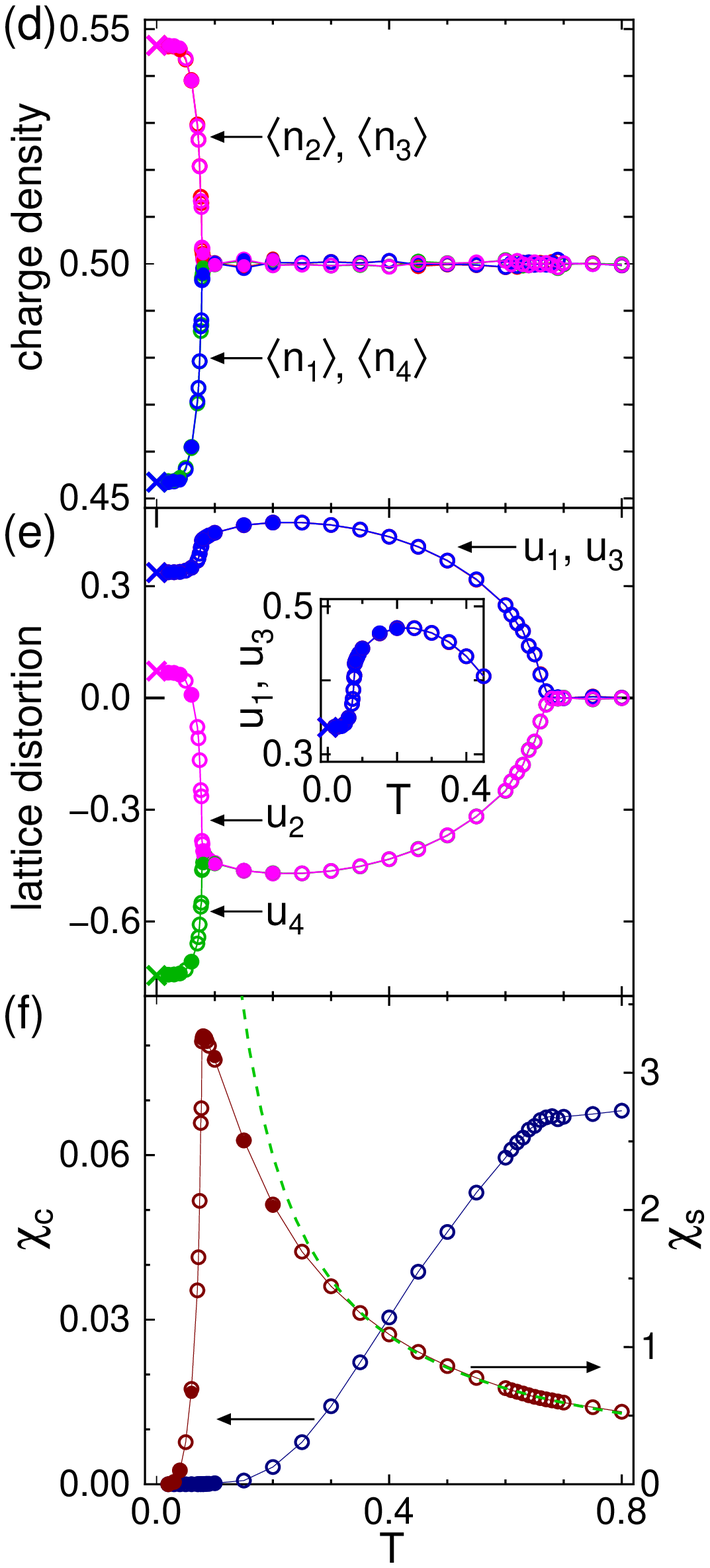}
  \caption{%(Color online): 
  Temperature dependences of
  charge 
  densities, 
  lattice distortions, and 
  charge and magnetic susceptibilities
  for
  $U=6$, $V=2.5$, $K_{\text{P}}=0.75$, $\delta_{\text{d}}=0$, and $K_{\text{P}_{2}}=0$;
  (a)-(c) $zV_{\perp}=1.5$,
  (d)-(f) $zV_{\perp}=0.25$.
  Open and closed circles represent results for $N$=32 and 64, respectively.
  Statistical errors are smaller than the symbol sizes. 
  Data shown by cross symbols at $T$=0 are 
  obtained by the exact diagonalization (Lanczos) method for $N$=12.
  The dashed line in (c) [(f)] shows a Curie-Weiss fit
  for $0.3 < T < T_{\text{CO}} (T_{\text{DM}})$.
  \label{fig:CO} 
  }
 \end{center}
\end{figure}

First, let us show the results
in the absence of intrinsic dimerization, i.e., $\delta_{\text{d}}=0$. 
Figure~\ref{fig:CO} shows 
the $T$-dependences of 
the charge densities $\langle n_{i} \rangle$,
the lattice distortions $u_i$, and 
the charge and magnetic susceptibilities $\chi_{\text{c}}$ and $\chi_{\text{s}}$,
in the case of the harmonic lattice ($K_{\text{P}_{2}}=0$).
Figures~\ref{fig:CO}(a)-\ref{fig:CO}(c) show the results for $zV_{\perp}=1.5$,
which are typical behaviors when the interchain Coulomb interaction is dominant
compared with the electron-lattice coupling, and 
Figs.~\ref{fig:CO}(d)-\ref{fig:CO}(f) show those for an opposite situation with $zV_{\perp}=0.25$.

In the former case, 
CO takes place at $T = T_{\text{CO}} \simeq 0.86$,
as observed in 
the alternation of $\langle n_{i} \rangle$ 
shown in Fig.~\ref{fig:CO}(a).
The charge susceptibility $\chi_{\text{c}}$ 
($= \partial n / \partial \mu$, $n$: average electron density)
suddenly drops at $T_{\text{CO}}$ and decreases rapidly 
below it owing to the opening of a charge gap,
while the magnetic susceptibility $\chi_{\text{s}}$ 
($= \partial m / \partial h$, 
 $m$: average magnetic moment, 
 $h$: magnetic field)
show no obvious change
at $T_{\text{CO}}$~\cite{QTM,Tanaka-Ogata-2005} [Fig.~\ref{fig:CO}(c)].
This is the transition to a paramagnetic insulating state.
The order parameter for CO develops down to $T\simeq 0.3$, 
below which, unexpectedly, it slightly decreases
[the inset of Fig.~\ref{fig:CO}(a)]. 
This temperature corresponds to the region
where $\chi_{\text{s}}$ deviates from the Curie-Weiss behavior
[Fig.~\ref{fig:CO}(c)], 
i.e., where spins start to interact with each other. 
Hence, we consider that 
this reduction of CO is due to 1D fluctuations 
fully taken into account in our scheme.
With further decrease in $T$, 
tetramerization emerges 
in both the charge and bond  sectors 
at $T = T_{\text{CO+SP}}\simeq 0.06$, as shown in Figs.~\ref{fig:CO}(a) and \ref{fig:CO}(b),
which was not accessible in the previous QTM results~\cite{QTM}. 
This transition is ascribed to the SP transition, 
where two neighboring spins at charge-rich sites form a spin-singlet pair
with spontaneous lattice distortions:
the low-$T$ phase is the CO+SP state~\cite{Kuwabara-Seo-Ogata-2003} 
sketched in Fig.~\ref{fig:phase_diagram}(d). 
The sudden drop in $\chi_{\text{s}}$ 
below $T_{\text{CO+SP}}$ in Fig.~\ref{fig:CO}(c)
confirms that the system has a spin gap.

Different successive transitions occur 
when the electron-lattice coupling is dominant
[Figs.~\ref{fig:CO}(d)-\ref{fig:CO}(f)].
In this case,
the lattice dimerization first takes place
at $T = T_{\text{DM}} \simeq 0.67$ 
[Fig.~\ref{fig:CO}(e)],
which leads the system effectively half-filled.
This indicates the DM transition with the opening of a charge gap,
as seen in the strong suppression of $\chi_{\text{c}}$ 
[Fig.~\ref{fig:CO}(f)].
$\chi_{\text{s}}$ shows no significant change at $T_{\text{DM}}$, 
and again this is characteristic of the transition to 
a paramagnetic insulator. 
Similarly to that in the CO case,
the DM order parameter decreases slightly below $T\simeq 0.2$ 
[the inset of Fig.~\ref{fig:CO}(e)],
where $\chi_{\text{s}}$ deviates from the Curie-Weiss behavior 
[Fig.~\ref{fig:CO}(f)]. 
Symmetry breaking with a four-site period occurs
at $T = T_{\text{DM+SP}} \simeq 0.08$, as shown in Figs.~\ref{fig:CO}(d) and \ref{fig:CO}(e).
This is another SP transition 
from the DM state to the DM+SP state~\cite{Ung-Mazumdar-Toussaint-1994} 
with a sudden drop in $\chi_{\text{s}}$ [Fig.~\ref{fig:CO}(f)], 
whose ordering pattern is different from that of 
the CO+SP state,
as shown in Fig.~\ref{fig:phase_diagram}(d).

Results obtained by varying $zV_\perp$ are summarized 
in the $zV_{\perp}$-$T$ phase diagram 
of Fig.~\ref{fig:phase_diagram}(a).
At high $T$,
there are phase transitions 
from a high-$T$ uniform metal 
to intermediate-$T$ paramagnetic insulators with 
the two-site period, 
the DM or CO state;
both transitions are of second order.
The DM and CO states are competing with each other 
and the transition between them appears to be of first order.
Thus, the phase diagram is likely of bicritical type
as a result of this competition. 
We will come back to this point later.
At lower $T$, the tetramerizations emerge upon 
both the CO and DM states,
and these SP transitions are of second order.
Note that the low-$T$ phases 
were not identified in the previous QTM study
because of numerical limitations~\cite{QTM}. 

\begin{figure}[htbp]
 \begin{center}
  \includegraphics[width=0.240\textwidth,clip]{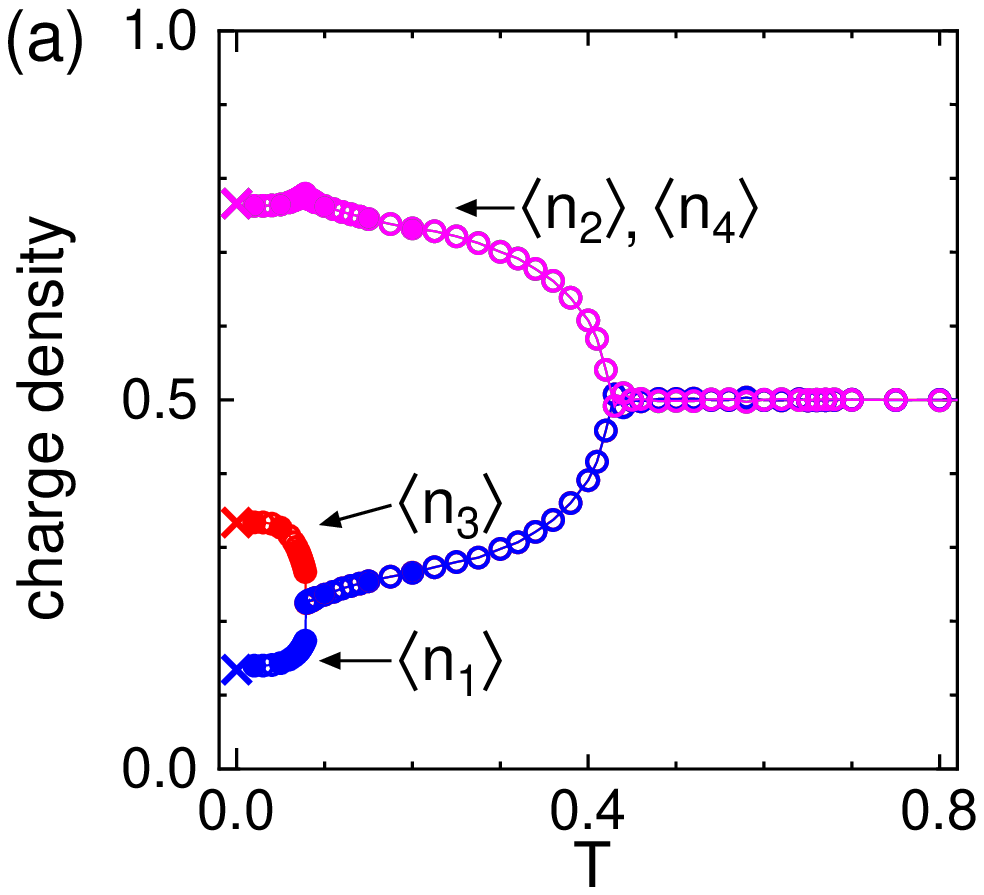}
  \includegraphics[width=0.240\textwidth,clip]{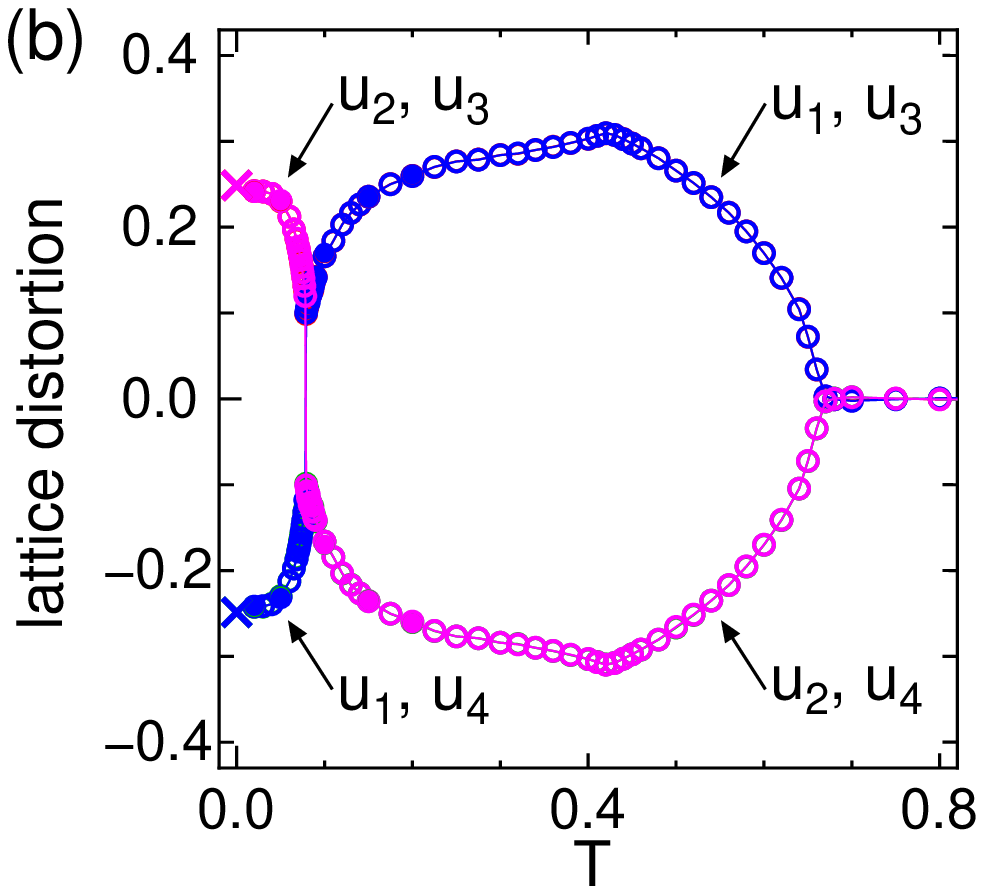}
  \caption{%(Color online): 
  Temperature dependences of
  (a) charge 
  densities
  and (b) lattice distortions 
  for
  $U=6$, $V=2.5$, $K_{\text{P}}=0.75$,
  $\delta_{\text{d}}=0$, $K_{\text{P}_{2}}=0.75$, and $zV_{\perp}=0.9$.
  The same symbols as those in Fig.~\ref{fig:CO} are used.
  \label{fig:Kp2} 
 }
 \end{center}
\end{figure}

The phase diagram is also investigated 
in the case of anharmonic lattice distortions,
as shown in Fig.~\ref{fig:phase_diagram}(b), for $K_{\text{P}_{2}} = 0.75$.
The anharmonicity opens a window 
where DM and CO coexist, and therefore,
the phase diagram shows a tetracritical behavior.
Although the coexisting phase was already found in ref.~\citen{QTM}, 
here we elucidate that 
the lattice anharmonicity is a controlling parameter of multicriticality.
It is clear that 
the anharmonicity $K_{\text{P}_{2}}$
stabilizes the coexistence of DM and CO. 
We note that the harmonic model with $K_{\text{P}_{2}}=0$ 
appears to be
on the verge of bicritical and
tetracritical behaviors~\cite{elsewhere},
even though it is rather difficult 
to exclude the coexistence in a very narrow range.
Typical $T$ dependences of the order parameters 
in the coexisting regime are shown in Fig.~\ref{fig:Kp2} for $zV_{\perp}=0.9$.
At $T_{\text{DM}}\simeq 0.67$, 
dimerization occurs first in the bond sector, and 
CO appears at a lower $T_{\text{DM+CO}} \simeq 0.42$;
CO coexists with DM below $T_{\text{DM+CO}}$.
With further decreasing $T$,
the degree of lattice dimerization decreases as CO grows, and finally,
the system exhibits a transition to the CO+SP state. 
This transition appears to be of first order
since the lattice dimerization suddenly vanishes
as shown in Fig.~\ref{fig:Kp2}(b).
For large or small $zV_\perp$, 
the coexistence of DM and CO disappears, and 
the successive transitions as in Fig.~\ref{fig:CO} take place. 
The results are summarized in Fig.~\ref{fig:phase_diagram}(b).

\begin{figure}[htbp]
 \begin{center}
  \includegraphics[width=0.240\textwidth,clip]{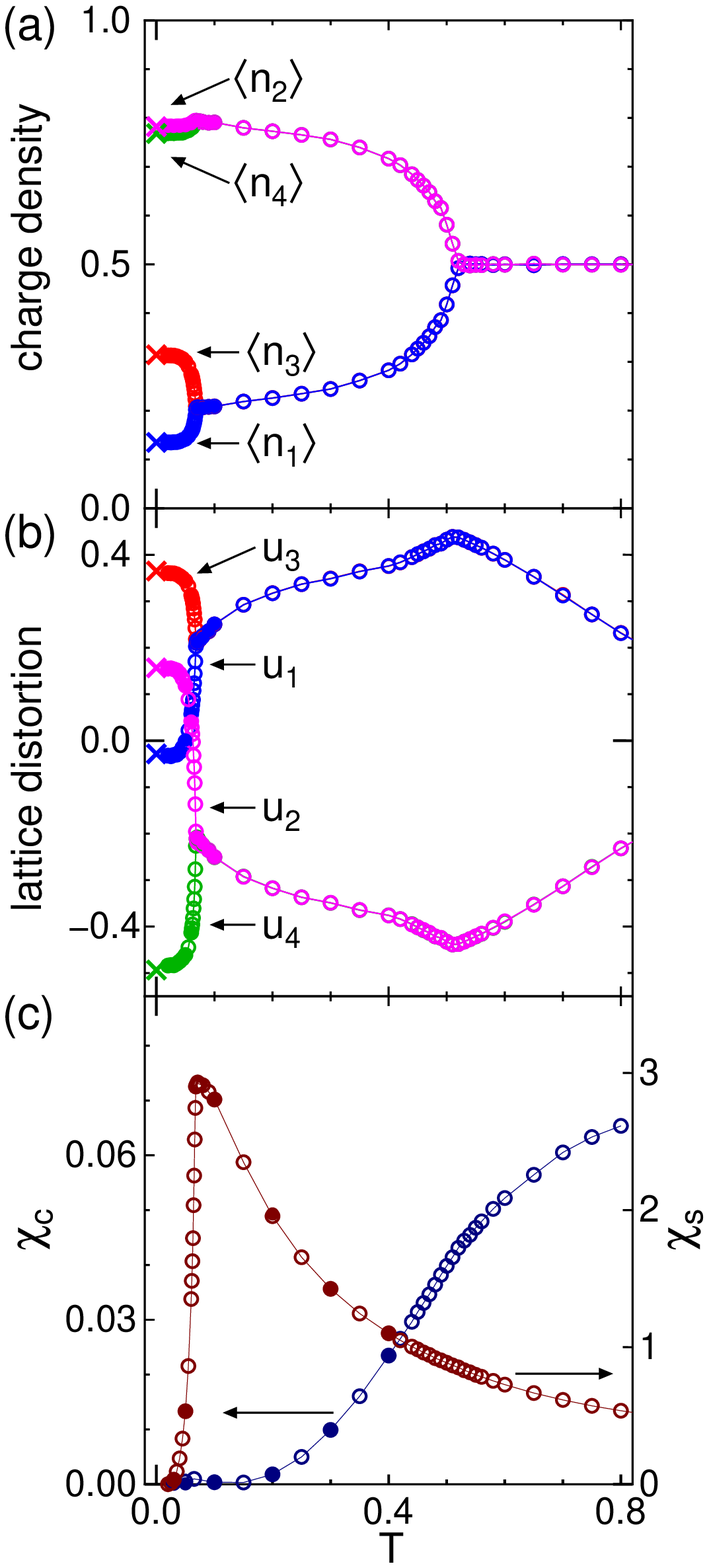}
  \includegraphics[width=0.240\textwidth,clip]{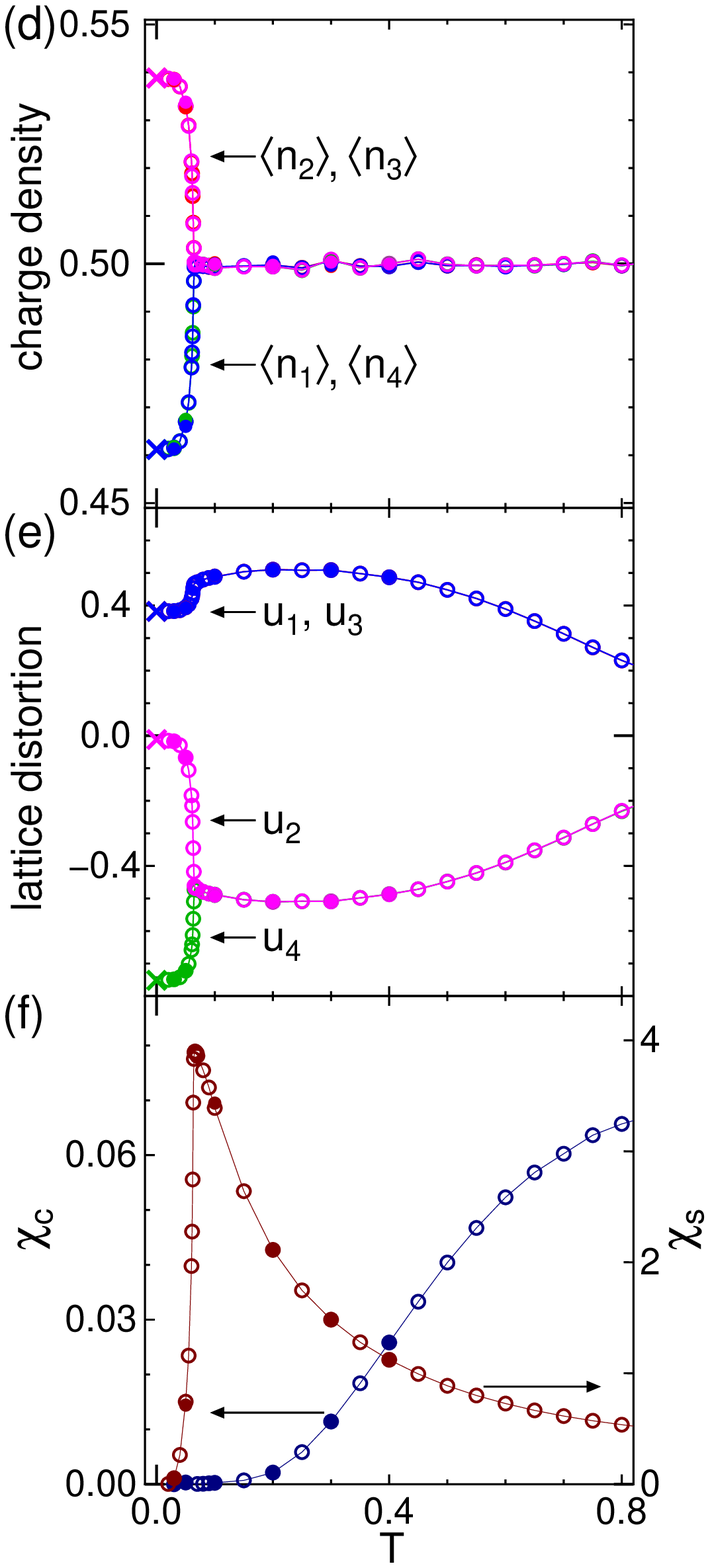}
  \caption{
  Temperature dependences of
  charge 
  densities, 
  lattice distortions, and 
  charge and magnetic susceptibilities
  for
  $U=6$, $V=2.5$, $K_{\text{P}}=0.75$, $\delta_{\text{d}}=0.02$, and $K_{\text{P}_{2}}=0$;
  (a)-(c) $zV_{\perp}=1.2$,
  (d)-(f) $zV_{\perp}=0.2$.
  The same symbols as those in Fig.~\ref{fig:CO} are used.
  \label{fig:DM} 
  }
 \end{center}
\end{figure}

Finally, we study the effects of 
the intrinsic lattice dimerization $\delta_{\text{d}}$.
The phase diagram for $\delta_{\text{d}}=0.02$ 
is represented in Fig.~\ref{fig:phase_diagram}(c)
and typical $T$ dependences of physical quantities 
are shown in Fig.~\ref{fig:DM}. 
For $\delta_{\text{d}} \neq 0$,
there is no DM transition 
since the lattice dimerization is always induced 
even at high $T$ [Figs.~\ref{fig:DM}(b) and \ref{fig:DM}(e)].
We thus refer to
the dimerized state without spontaneous symmetry breaking at high $T$
as the intrinsic DM (iDM) state
in the phase diagram in Fig.~\ref{fig:phase_diagram}(c).
When $zV_{\perp}$ is dominant, 
CO and SP transitions occur similarly
to the cases of $\delta_{\text{d}} = 0$ as shown in Figs.~\ref{fig:DM}(a) and \ref{fig:DM}(b). 
Note, however, that
the CO+SP(+iDM) state here is slightly different from 
the CO+SP state for $\delta_{\text{d}}=0$ 
because of the underlying lattice dimerization [Fig.~\ref{fig:phase_diagram}(d)].
On the other hand, when the electron-lattice coupling is dominant,
CO is suppressed and the iDM state prevails 
until the tetramerization takes place at lower $T$ [Figs.~\ref{fig:DM}(d) and \ref{fig:DM}(e)].
$\chi_{\text{s}}$ behaves similarly to that for $\delta_{\text{d}}=0$
in both cases
[Figs.~\ref{fig:DM}(c) and \ref{fig:DM}(f)],
indicating that the intrinsic lattice dimerization 
have no significant effect on the spin sector.
On the other hand, $\chi_{\text{c}}$ behaves differently at high $T$:
it is suppressed even at $T > T_{\text{CO}}$ 
because of the Mott insulating nature of the iDM state.
The phase diagram is summarized in Fig.~\ref{fig:phase_diagram}(c), 
where the DM transition is absent and the phase transition between 
the iDM and CO(+iDM) phases is of second order. 

%\texttt{[Discussion]}

Now let us discuss our results in comparison with the experimental data.
The family of $(R_{1}R_{2}$-DCNQI)$_{2}X$ has uniform DCNQI chains at high $T$, 
corresponding to the cases of $\delta_{\text{d}}=0$ in our model. 
The successive transitions with dimerization and tetramerization 
observed in DMe-DCNQI$_2$Ag \cite{Moret-1988}
are reproduced by the DM and SP transitions,
as shown in Figs.~\ref{fig:CO}(d)-\ref{fig:CO}(f). 
The CO transition in DI-DCNQI$_2$Ag~\cite{Hiraki-Kanoda-1998} 
is also reproduced in Figs.~\ref{fig:CO}(a)-\ref{fig:CO}(c), 
but the low-$T$ antiferromagnetism~\cite{Hiraki-Kanoda-1996} 
is out of the scope of our model
because we neglect the interchain hopping 
that results in
an effective antiferromagnetic exchange interaction
between chains~\cite{Riera}.
On the other hand, 
(TMTTF)$_{2}X$ has an intrinsic dimerization, 
and hence corresponds to $\delta_{d} \neq 0$. 
The CO and SP transitions observed for $X$=PF$_{6}$ and AsF$_{6}$
are reproduced in Figs.~\ref{fig:DM}(a)-\ref{fig:DM}(c).
Furthermore, our phase diagram in Fig.~\ref{fig:phase_diagram}(c) 
shows a good agreement with the $P$-$T$ phase diagram 
for $X$=AsF$_{6}$~\cite{Zamborszky}
when we read the interchain Coulomb 
interaction $zV_\perp$ as the inverse of $P$.
This is reasonable since 
the pressure is supposed to increase mainly the intrachain transfer integrals, 
i.e., to effectively decrease the interchain interaction.
As to the pressure effect on DCNQI compounds, 
the case for DMe-DCNQI$_2$Ag has not been explored yet, but
it was shown that DI-DCNQI$_2$Ag exhibits 
a peculiar tricritical behavior under $P$~\cite{Itoh}.
Although we have found no tricritical point
in Figs.~\ref{fig:phase_diagram}(a) and \ref{fig:phase_diagram}(b),
our observation of the role of lattice anharmonicity in the multicriticality
indicates the possibility of capturing such behavior 
by extending our model, particularly in the electron-lattice part.

It should be pointed out that
the obtained phase diagrams will be modified
if we go beyond 
the interchain mean-field or adiabatic approximation.
For example, 
the effects of fluctuations and quantum nature in the lattice
are anticipated to contract the DM and SP phases.
These phases would nevertheless survive at finite $T$
provided that 
three-dimensional lattice couplings are included. 
Hence, we presume that 
the general features of our phase diagrams remain robust, 
and bear the comparisons with experiments.
The effects of fluctuations neglected in the present calculations 
are left for future study.

%\texttt{[Summary]}

In summary,
we have investigated the finite-temperature phase transitions
in quasi-one-dimensional quarter-filled systems
using the extended Hubbard model with electron-lattice coupling.
The effective one-dimensional model, 
obtained by interchain mean-field approximation,
has been numerically solved by the stochastic-series-expansion Monte Carlo method
down to temperatures far below that previously studied.
A variety of phase diagrams have been explored 
and rich behaviors of competition and coexistence
have been clarified among 
charge, lattice and spin degrees of freedom.
We have found that 
an anharmonicity of lattice distortions is  
a key parameter of multicritical behaviors. 
Our results reproduce well 
the charge ordering, dimer Mott, and spin-Peierls states
in DCNQI and TMTTF compounds.

%\texttt{[Acknowledgment]}
\section*{Acknowledgments}

The authors thank 
S. E. Brown, 
R. T. Clay,
S. Fujiyama, 
H. Yoshioka, 
and M. Tsuchiizu for fruitful discussions. 
% YO appreciates T. Kato for encouraging comments.
This work is supported by Grants-in-Aid for
Scientific Research 
(Nos.
18028018, 
18028026, 
19014020)
from the Ministry of Education, Culture, Sports, Science and Technology, and
by %Next Generation Super Computing Project, Nanoscience Program.
Next Generation Integrated Nanoscience Simulation Software.

%\texttt{[References]} 
%\texttt{sort by order of appearance!}


\begin{thebibliography}{99}
\bibitem{elsewhere}
	We also found 
	that the anharmonicity changes
	the multicritical behaviors between tetramized SP phases, 
	which will be discussed elsewhere.
\end{thebibliography}
\end{document}